\documentclass[twocolumn,amsmath,amssymb,aps,longbibliography,floatfix,8pt]{revtex4-1}

\usepackage{cancel}
\usepackage{enumitem}
\usepackage[utf8x]{inputenc}
\usepackage{graphicx}
\usepackage{dcolumn}
\usepackage{bm}
\usepackage{hyperref}
\usepackage[switch]{lineno}
\usepackage{float}             
\usepackage{subfigure}
\usepackage{tikz}
\usetikzlibrary{arrows,decorations.pathmorphing,backgrounds,positioning,fit,petri}

\begin{document}

\title{Multiple Spin-Phonon Relaxation Pathways in a Kramer Single-Ion Magnet}

\author{$^{1}$Alessandro Lunghi}
\email{lunghia@tcd.ie}
\author{$^{1}$Stefano Sanvito}
\affiliation{$^{1}$School of Physics, CRANN Institute and AMBER, Trinity College, Dublin 2, Ireland}

\begin{abstract}
{\bf We present a first-principles investigation of spin-phonon relaxation in a molecular crystal of Co$^{2+}$ single-ion magnets. Our study combines electronic structure calculations with machine-learning force fields and unravels the nature of both the Orbach and the Raman relaxation channels in terms of atomistic processes. We find that although both mechanisms are mediated by the excited spin states, the low temperature spin dynamics is dominated by phonons in the THz energy range, which partially suppress the benefit of having a large magnetic anisotropy. This study also determines the importance of intra-molecular motions for both the relaxation mechanisms and paves the way to the rational design of a new generation of single-ion magnets with tailored spin-phonon coupling.}
\end{abstract}

\maketitle

\section*{Introduction}

Storing information at the molecular level is a long-standing dream of chemists and material scientists~\cite{Sessoli1993}. In magnetic information storage, single-ion magnets (SIMs) based on Kramer ions~\cite{Zadrozny2011} have received extensive attention due to the possibility to synthesize coordination compounds with large easy-axis magnetic anisotropy and almost-perfect axial symmetry~\cite{Goodwin2017,Bunting2018}. Under these conditions the molecular magnetic moment is stabilized against temperature, since the spin relaxation must occur through the interaction with the lattice vibrations and requires an energy exchange proportional to the size of the molecular magnetic anisotropy, 
$U_\mathrm{eff}$. This relaxation regime is named after Orbach. The relaxation time generally follows an Arrhenius-like law with temperature, $\tau=\tau_{0}exp(U_\mathrm{eff}/k_\mathrm{B}T)$, where $k_\mathrm{B}$ is the Boltzman constant, $\tau_0$ is the attempt time and $T$ the temperature.

The design of the crystal field of SIMs has been, for decades, the driving strategy to increase $U_\mathrm{eff}$~\cite{Rinehart2011,Craig2015,Liddle2015,Frost2016,Harriman2019}. Such strategy has recently reached an important milestone with the synthesis of magnetic molecules having a blocking temperature above the nitrogen's boiling point~\cite{Guo2018}. Despite these accomplishments, the improvement of the spin lifetime of these complexes has not completely matched the expectations raised by designing magnetic anisotropy barriers above 1000~$K$. This is mostly due to the fact that i) the attempt time $\tau_{0}$ takes very small values~\cite{Rajnak2016,Giansiracusa2019} and ii) an additional relaxation mechanism, often interpreted as Raman relaxation, appears at low temperature~\cite{Harriman2019}. The two effects pose strong limitations 
to the progress in this field and, despite their crucial importance, they are still not completely understood 
at the atomistic level. 

Here we provide a microscopic explanation of these phenomena by delivering a comprehensive first-principles description of the spin relaxation of a prototypical SIM with large axial magnetic anisotropy, namely [Co(L)$_{2}$]$^{2-}$, where H$_{2}$L=1,2-bis(methanesulfonamido)benzene\cite{Rechkemmer2016}. Our method is based on state-of-the-art electronic structure theory~\cite{Neese2019}, first-principles spin relaxation~\cite{Lunghi2019b} and machine-learning tools~\cite{Lunghi2019} and it makes it possible to deliver the a parameter-free description of both one- and two-phonon processes leading to spin relaxation in SIMs, with the inclusion of all the phonons in the Brillouin zone. 

\section*{Results and Discussion}

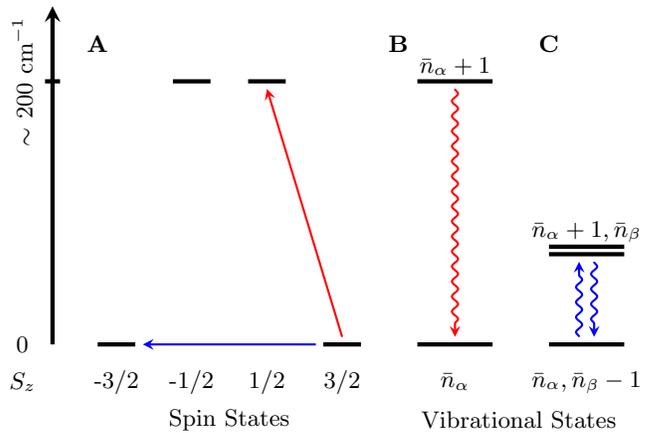
\begin{figure}
\begin{center}
\begin{tikzpicture}

\node at (-4,6) {\textbf{A}};
\node at (0,6) {\textbf{B}};
\node at (2,6) {\textbf{C}};

\draw [ultra thick, ->,>=stealth] (-4.6,2) -- (-4.6,6.5);
\node [rotate=90] at (-5,5.5) {$\sim$ 200 cm$^{-1}$};
\draw [ultra thick] (-4.5,5.5) -- (-4.7,5.5);
\node at (-5,2) {0};
\node at (-5,1.5) {$S_{z}$};
\draw [ultra thick] (-4,2) -- (-3.5,2);
\draw [ultra thick] (-3,5.5) -- (-2.5,5.5);
\draw [ultra thick] (-2,5.5) -- (-1.5,5.5);
\draw [ultra thick] (-1,2) -- (-0.5,2);

\node at (-3.75,1.5) {-3/2};
\node at (-0.75,1.5) {3/2};
\node at (-2.75,1.5) {-1/2};
\node at (-1.75,1.5) {1/2};

\node at (-2.25,1) {Spin States};

\node at (1.6,1) {Vibrational States};

\draw [thick, red,->,>=stealth] (-0.75,2.1) -- (-1.75,5.4);
\draw [thick, blue,->,>=stealth] (-1.1,2) -- (-3.4,2);

\draw [ultra thick] (0.25,5.5) -- (1.25,5.5);
\draw [ultra thick] (0.25,2) -- (1.25,2); 

\draw [thick,red,->,>=stealth,decorate,
decoration={snake,amplitude=.4mm,segment length=2mm,post length=1mm}]
(0.75,5.4) -- (0.75,2.1);

\node at (0.75,1.5) {$\bar{n}_{\alpha}$};
\node at (0.75,5.7) {$\bar{n}_{\alpha}+1$};

\draw [ultra thick] (2,2) -- (3,2);


\draw [ultra thick] (2,3.3) -- (3,3.3);
\draw [ultra thick] (2,3.2) -- (3,3.2); 

\draw [thick,blue,->,>=stealth,decorate,
decoration={snake,amplitude=.4mm,segment length=2mm,post length=1mm}]
(2.4,2.1) -- (2.4,3.1) ;
\draw [thick,blue,->,>=stealth,decorate,
decoration={snake,amplitude=.4mm,segment length=2mm,post length=1mm}]
(2.6,3.1) -- (2.6,2.1) ;  

\node at (2.5,1.5) {$\bar{n}_{\alpha},\bar{n}_{\beta}-1$};
\node at (2.5,3.5) {$\bar{n}_{\alpha}+1,\bar{n}_{\beta}$};

\end{tikzpicture}
\end{center}
\caption{\textbf{Schematic Representation of the Orbach and Raman Relaxation Mechanisms.} 
(\textbf{A}) The spin states of Co(pdms)$_{2}$ are reported as function of their energy and the 
nominal S$_{z}$ value. The red and blue arrows represent possible relaxation pathways. 
(\textbf{B}) An electronic excitation from the ground state to an excited state (red arrow in panel 
\textbf{A}) can be accompanied by the absorption of a phonon with energy in resonance with the 
spin transition (Orbach process). (\textbf{C}) The direct transition S=3/2 $\rightarrow$ S=-3/2 
(blue arrow in panel \textbf{A}) can be induced by the simultaneous absorption and emission of 
two phonons with the same energy (Raman process).}
\label{scheme}
\end{figure}

The spin properties of transition-metal-based SIMs, such as our Co$^{2+}$ complex, are described by a spin Hamiltonian including a single-ion anisotropy term
\begin{equation}
\hat{H}_\mathrm{S}=\vec{\mathbf{S}}\cdot \mathbf{D} \cdot \vec{\mathbf{S}}\:,
\label{SH}
\end{equation}
where $\vec{\mathbf{S}}$ describes the $|S|=3/2$ ground state of a high-spin Co$^{2+}$. The calculated 
anisotropy tensor, $\mathbf{D}$, correctly captures the easy-axis nature of the complex and predicts an 
excited Kramer doublet at $\sim 200$ cm$^{-1}$. Fig.~\ref{scheme}A offers a graphical representation of 
the spin states. In comparison, experiments place the Kramer excited state at $230$ cm$^{-1}$ above 
the ground state~\cite{Rechkemmer2016}.

The spin Hamiltonian of Eq.~(\ref{SH}) is the starting point for the treatment of the spin relaxation. The 
spin-phonon coupling interaction is extracted from the dependence of the anisotropy tensor $\mathbf{D}$ 
on the atomic positions~\cite{Lunghi2017a}. At the first order in the coupling strength this interaction is 
described by the Hamiltonian 
\begin{equation}
\hat{H}_\mathrm{S-ph}=\sum_{\alpha\mathbf{q}} \hat{V}_{\alpha\mathbf{q}}\hat{Q}_{\alpha\mathbf{q}}=\sum_{\alpha\mathbf{q}}\left[\vec{\mathbf{S}}\cdot\left(\frac{\partial \mathbf{D}}{\partial Q_{\alpha\mathbf{q}}}\right)\cdot \vec{\mathbf{S}}\right]\hat{Q}_{\alpha\mathbf{q}}\:,
\label{SPH}
\end{equation}
where $\hat{Q}_{\alpha\mathbf{q}}$ represents the $\alpha$-phonon with reciprocal vector $\mathbf{q}$ 
and vibrational frequency $\hbar\omega_{\alpha\mathbf{q}}$. Within this weak-coupling regime, here we 
consider time-dependent processes at both the first and second order of  time-dependent perturbation theory.

At the first-order the transition rate between two spin states  $| a \rangle$ and $| b \rangle$ is provided by the 
expression
\begin{equation}
W^\mathrm{1-ph}_{ba} = \frac{2\pi}{\hbar^{2}} \sum_{\alpha\mathbf{q}} \big|\langle b | \hat{V}_{\alpha\mathbf{q}} | a \rangle \big|^{2} G^\mathrm{1-ph}(\omega_{ba},\omega_{\alpha\mathbf{q}}) \:,
\label{order1}
\end{equation}
where $G^\mathrm{1-ph}=\delta(\omega-\omega_{\alpha\mathbf{q}})\bar{n}_{\alpha\mathbf{q}}+\delta(\omega+\omega_{\alpha\mathbf{q}})(\bar{n}_{\alpha\mathbf{q}}+1)$ and 
$\bar{n}_{\alpha\mathbf{q}}=[exp(\hbar\omega_{\alpha}/k_\mathrm{B}T)-1]^{-1}$ is the Bose-Einstein 
thermal population. The first (second) term of $G^\mathrm{1-ph}$ accounts for a spin transition due to the 
absorption (emission) of a single phonon. For perfectly axial Kramer systems in the absence of external 
interactions able to break time-reversal symmetry, the direct transition $S_{z}=3/2$ $\rightarrow$ 
$S_{z}=-3/2$ is forbidden and relaxation based on Eq.~(\ref{order1}) must proceed through the excited 
state by absorption of a phonon. Fig.~\ref{scheme}B schematically reports such a process.

At the second order two phonons can simultaneously mediate the spin relaxation, if an intermediate spin 
state, $| c \rangle$, allows the transition. In this case the rate of population transfer between two spin states 
$| a \rangle$ and $| b \rangle$ is given by
\begin{align}
W^\mathrm{2-ph}_{ba} = \frac{2\pi}{\hbar^{2}} \sum_{\alpha\mathbf{q}}\sum_{\beta\mathbf{q'}} & \Big|\sum_{c}\frac{\langle b |\hat{V}_{\alpha\mathbf{q}}|c\rangle\langle c|\hat{V}_{\beta\mathbf{q'}}| a \rangle}{E_{c}-E_{a}\pm\hbar\omega_{\beta\mathbf{q'}}}\Big|^{2}\cdot \\
& \cdot G_{\pm}^\mathrm{2-ph}(\omega_{ba},\omega_{\alpha\mathbf{q}},\omega_{\beta\mathbf{q'}}) \:,
\label{order2}
\end{align}
where $G_{\pm}^\mathrm{2-ph}$, similarly to $G^\mathrm{1-ph}$ for single phonon processes, accounts for the
energy conservation and the phonon thermal population of two-phonon interactions. For instance, a simultaneous 
absorption and emission of two phonons with virtually the same energy, schematically described in Fig.~\ref{scheme}C, 
would contribute to the $T$ dependence of $W_{ba}^\mathrm{2-ph}$ through $G_{-}^\mathrm{2-ph}=\delta(\omega+\omega_{\alpha\mathbf{q}}-\omega_{\beta\mathbf{q'}})(\bar{n}_{\alpha\mathbf{q}}+1)\bar{n}_{\beta\mathbf{q'}}$. 
The complete expression for $G_{\pm}^\mathrm{2-ph}$ is provided in the Supporting Information (SI). 

\begin{figure}
 \begin{center}
  \includegraphics[scale=1]{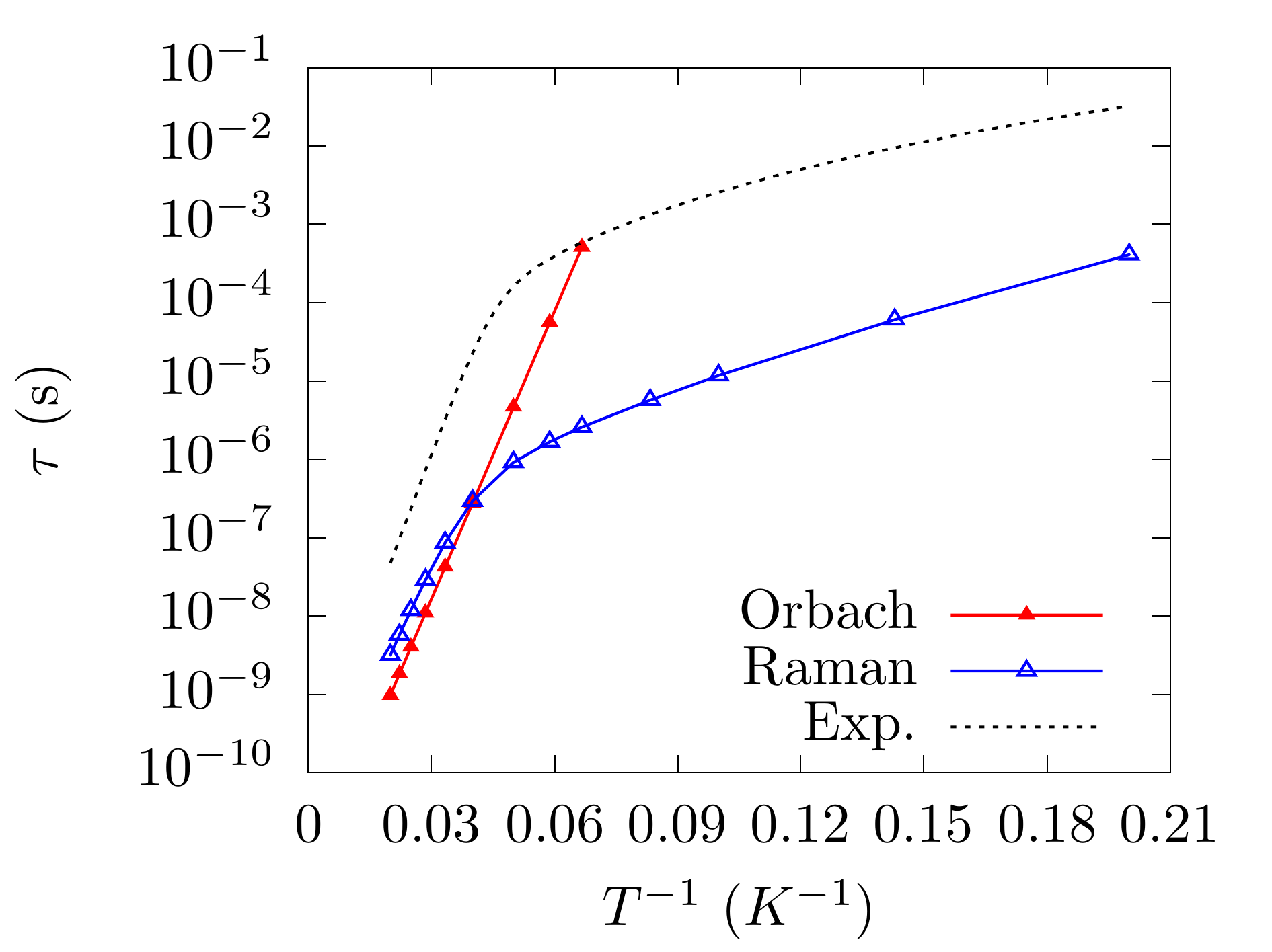}
  \includegraphics[scale=1]{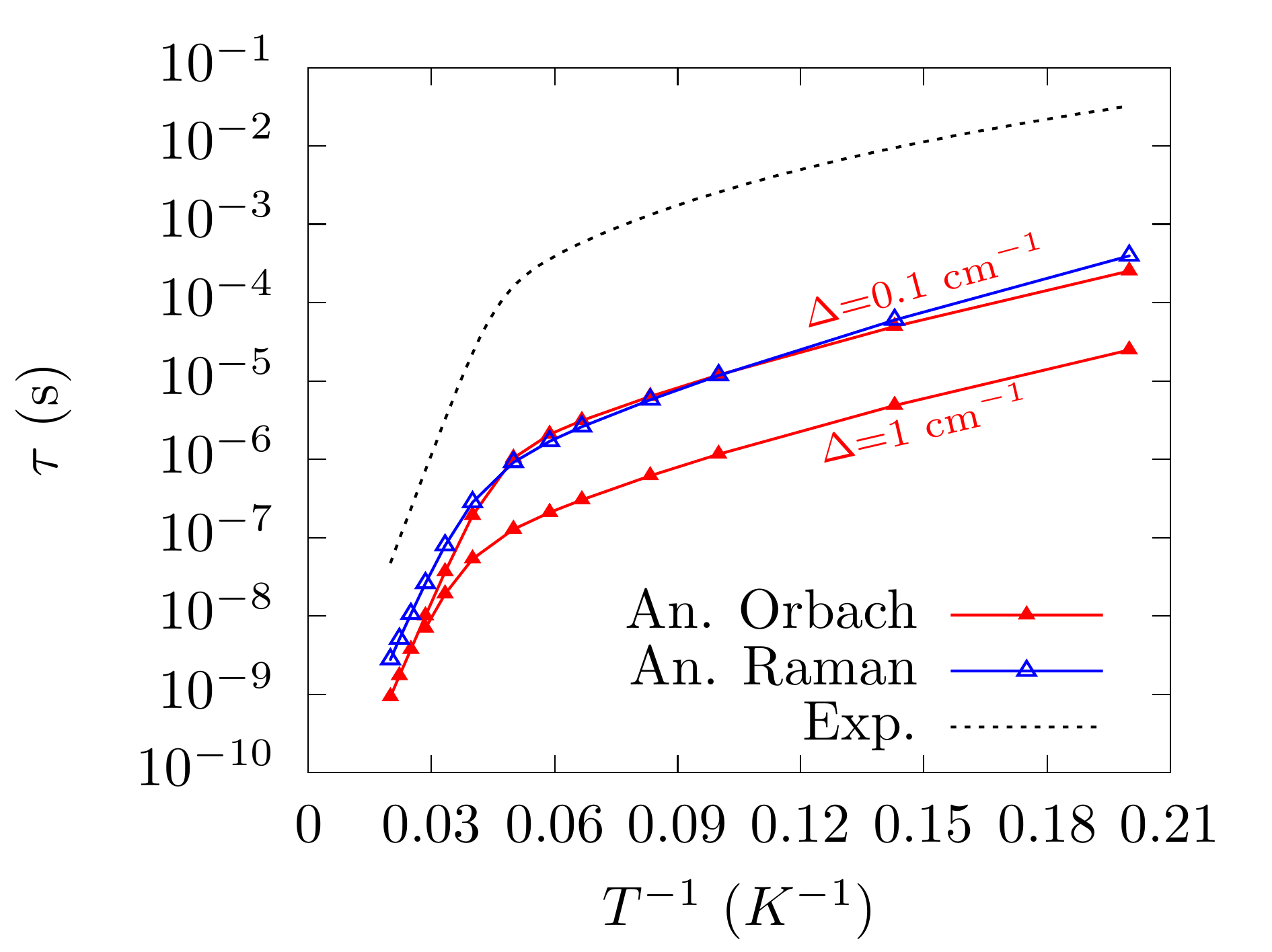}
 \end{center}
\caption{\textbf{Temperature Dependence of the Spin-Phonon Relaxation Times.} 
The top (bottom) panel reports the calculated relaxation time as function of temperature for the case 
of harmonic (anharmonic) phonons. One-phonon processes are reported with a continuous red line 
and filled triangles, while two-phonon processes are reported with continuous blue line and empty 
triangles. The dashed black line represents the best fit to the experimental data~\cite{Rechkemmer2016} 
obtained with the expression $\tau^{-1}=CT^{-n}+\tau_{0}^{-1}exp(-U_{eff}/K_{B}T)$, where $C=0.087$, 
$n=3.65$, $\tau_{0}=1.31$ $10^{-10}$ $s^{-1}$ and $U_{eff}=230$ cm$^{-1}$.}
 \label{tau}
\end{figure}

\begin{figure*}
\begin{center}
 \includegraphics[scale=1]{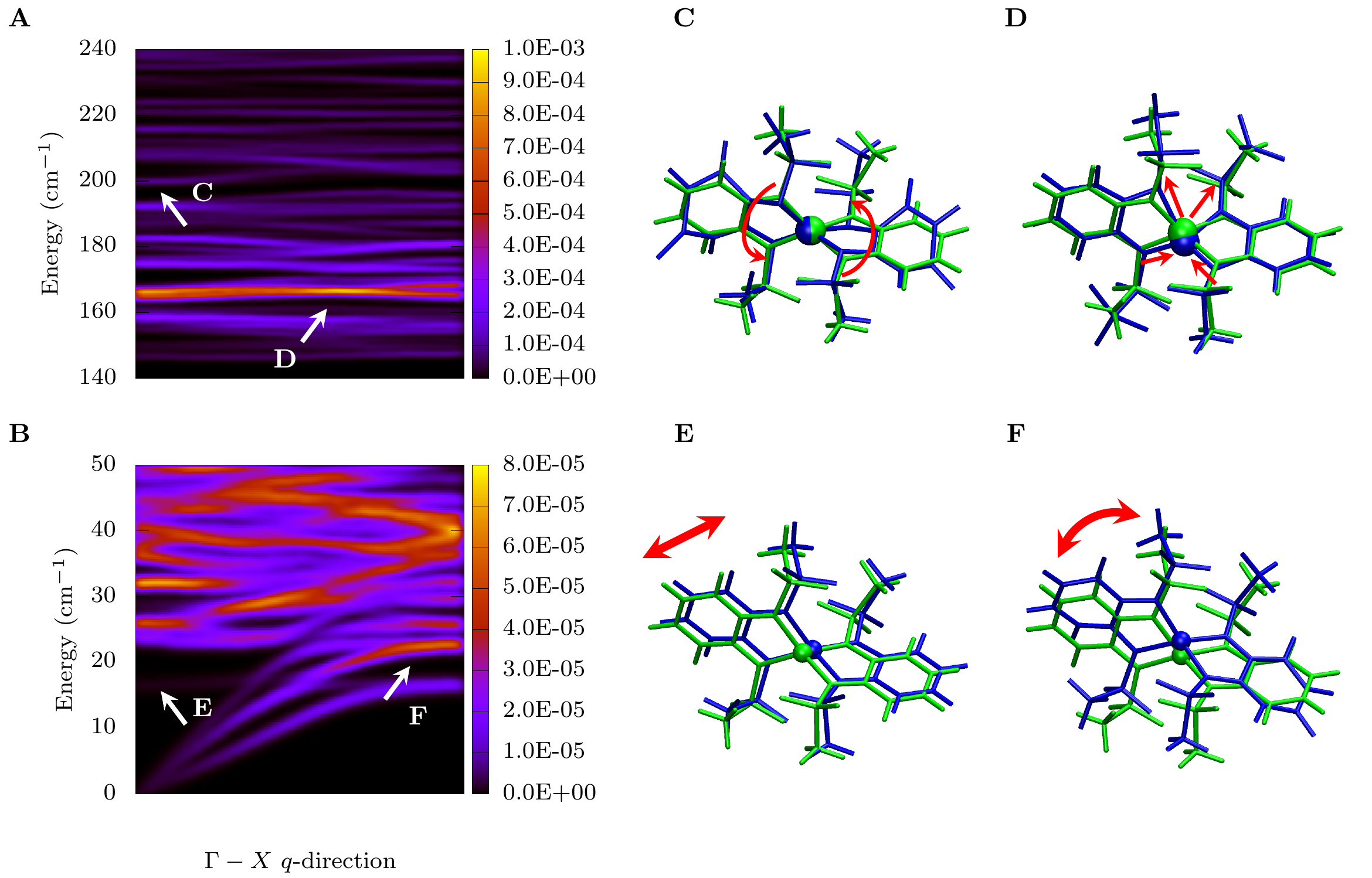}
\end{center}
\caption{\textbf{Mode-Resolved Spin-Phonon Coupling Intensity.} The spin-phonon coupling intensity, 
expressed in (cm$^{-1}/$\AA)$^{2}$ units, is reported for every mode along the $\Gamma=(0,0,0)-X=(0.5,0,0)$ 
path in the Brillouin zone over the energy window $140-240$ cm$^{-1}$ (\textbf{A}) and the energy window 
$0-50$ cm$^{-1}$ (\textbf{B}). The white arrows and labels in Panels \textbf{A} and \textbf{B} are associated 
with the corresponding molecular distortions in panels \textbf{C} to \textbf{F}, where the equilibrium molecular 
structure is represented in green and the distorted one in blue. Red arrows highlight the most important component 
of the molecular distortion. For simplicity only the magnetic ion is represented as a sphere.}
 \label{modes}
\end{figure*}

Here we implement Eqs.~(\ref{order1}) and (\ref{order2}) in a first-principles computational scheme able to 
provide an estimate of the spin relaxation time. The numerical calculation of the spin relaxation time, either 
at the first or second order of perturbation theory, requires the calculation of the single-ion tensor derivatives, 
$(\partial \mathbf{D}/\partial Q_{\alpha\mathbf{q}})$, of the phonons's frequencies, $\hbar\omega_{\alpha\mathbf{q}}$, 
and of the normal modes' compositions, $\hat{Q}_{\alpha\mathbf{q}}$, across the entire Brillouin zone. The 
numerical calculation of the first-order derivatives of the molecular anisotropy can be performed relatively easily 
within a first-principles scheme, as discussed before~\cite{Lunghi2017a}, and it requires $\sim 1000$ CASSCF 
simulations. Phonons calculations represent a bigger challenge. SIMs generally comprise tens of atoms and 
often crystallize together with counter ions in a quite large unit cell. This is also the case for our [CoL$_{2}$]$^{2-}$, 
whose unit cell contains 396 atoms. 

The ab-initio calculation of the phonons at a generic point in the Brillouin zone requires the use of multiples of 
the unit cell and, except for simple systems~\cite{Garlatti2020a}, it is generally impractical. Here we solve this 
challenge by employing a machine-learning force field approach. The model is based on the spectral neighbour 
analysis potential (SNAP)~\cite{Lunghi2019} augmented with van der Walls and electrostatic interactions. The 
latter are based on a Lennard Jones potential and atomic point charges, respectively. After the successful training 
of the model on just $\sim 200$ DFT reference calculations, the force field is used to optimize a 
$2\times2\times2$ super-cell containing 3168 atoms and calculate all the lattice harmonic force constants. The 
machine-learning force field is able to reproduce atomic forces within 3.0~kcal/mol/\AA$ $ from the DFT reference 
value. In comparison the training set contains atomic forces that span a 1200~kcal/mol/\AA$ $ range. A 
$\Gamma$-point DFT calculation of phonons, requiring 2375 DFT runs, is also performed in order to validate the 
ML model. An absolute mean error of just $\sim 6$~cm$^{-1}$ is observed for vibrations in the energy window of 
interest. More details on the model validation are provided in the SI. It should also be noted that the training of the 
ML force field requires a fraction of the computational cost needed to evaluate the sole $\Gamma$-point phonons 
with DFT and massively undercuts the cost of computing phonon dispersions by DFT.

The lattice force constants and the spin-phonon coupling coefficients, $(\partial \mathbf{D}/\partial Q_{\alpha\mathbf{q}})$, 
are used together with Eqs.~(\ref{order1}) and (\ref{order2}) to compute the spin-relaxation time as 
discussed in the SI. Results for both first- and second-order perturbation theory are reported in the top panel 
of Fig.~\ref{tau}, together with the best fit to experiments~\cite{Rechkemmer2016}. The first-order theory produces 
the expected exponential dependence of the spin relaxation with $T$. In contrast, Raman relaxation 
shows two different regimes: one pseudo-exponential that dominates $\tau$ below 30~$K$ and a second one 
that almost identically mimics Orbach relaxation at high $T$. 

We now proceed to discuss the potential role of lattice anharmonic interactions. Following the basic concepts of 
reference~\cite{Lunghi2017}, a finite phonon lifetime can be accounted for by substituting the Dirac functions 
appearing in $G^\mathrm{1-ph}$ and $G^\mathrm{2-ph}$ with a Lorentzian function. The predicted spin lifetimes 
obtained with a constant linewidth $\Delta$ in the range 0.1--1.0~cm$^{-1}$ are reported in the bottom panel of 
Fig.~\ref{tau}. The introduction of a finite phonon lifetime only affects the low-temperature regime of the Orbach 
relaxation mechanism, where $\tau$ assumes a $T$ profile virtually identical to the Raman one, with the only 
difference of being linearly dependent on the parameter $\Delta$. The results for the low $T$ regime, regardless 
the mechanism considered, are consistent with the commonly observed exponential behaviour, where a polynomial 
expression is used to model the relaxation time as function of temperature, \textit{i.e.} $\tau\propto T^{-n}$, with $n$ 
often in the range $2<n<4$~\cite{Liddle2015,Giansiracusa2019}. Although the computational results nicely reproduce 
the experimental profile, the overall $\tau$ is underestimated by $\sim 2$ orders of magnitude. This discrepancy is
attributed to numerical precision. Note that a careful benchmarking and an improvement of the computational methods 
for obtaining $\hat{V}_{\alpha\mathbf{q}}$ and the phonon-related quantities represent our future methodological 
challenge.

Let us now discuss the $\tau(T)$ profile in terms of atomistic processes. The slope of $ln(\tau)$ vs $T^{-1}$ in the 
Orbach regime, for $T>30K$, provides a $U_\mathrm{eff} \sim 200$ cm$^{-1}$. This corresponds to the absorption 
of phonons with energy in resonance to the excited Kramer spin doublet. For a perfectly harmonic phonon bath, 
this relation is preserved at any temperature. However, the inclusion of anharmonic interactions makes the absorption 
of out-of-resonance low-energy phonons the favourable single-phonon process at low temperature ($T<30K$), where 
the in-resonance phonons are too scarce~\cite{Lunghi2017}. Considering the Raman process, the transfer of population 
within the ground Kramer doublet state is driven by the simultaneous absorption and emission of two phonons with 
virtually the same energy, as depicted in Fig.~\ref{scheme}\textbf{C}. This interaction is made possible by the presence 
of spin excited states that mediate the interaction. From the dominator of Eq.~(\ref{order2}), it is clear that the phonons 
do not need to be in resonance with the spin excited states. For $T<30K$, phonons from the low energy window 
($\hbar\omega_{\alpha\mathbf{q}}=15-50$ cm$^{-1}$) provide the main contribution to spin relaxation and determine 
the $\tau(T)$ profile through $\bar{n}_{\alpha\mathbf{q}}(\bar{n}_{\alpha\mathbf{q}}+1)\sim \bar{n}_{\alpha\mathbf{q}}$. 

The deviation from a perfect exponential regime and the appearance of the polynomial behaviour is understood by 
considering that by increasing temperature more phonons become populated and start contributing to the spin relaxation. 
For $T>30K$, phonons with similar energy to the excited spin states becomes sufficiently populated and a two-phonon 
relaxation through the excited Kramer doublet becomes the dominant pathway. This process is initiated by the absorption 
of a phonon with energy just above the energy of the excited spin states and the emission of a second one that 
releases the excess of energy. This interpretation of Raman spin relaxation is also consistent with a recent 
experimental and theoretical investigation of a Lanthanide SIM~\cite{Chiesa2020}. It is interesting to note that 
Raman relaxation, receiving contributions from phonons across the entire Brillouin zone, is always able to fulfil 
the in-resonance condition with the spin states. For this reason it is not influenced by the inclusion of anharmonic 
interactions that introduce out-resonance absorption/emission. 

An important result of this study is that low-energy phonons, regardless of whether contributing through 
the anharmonic Orbach or the Raman mechanism, are always able to induce spin relaxation and limit the 
spin lifetime even at very low temperatures. In this regime, relaxation time still depends on the energy of 
the excited states, but through the relation $\tau\sim 4|D_{zz}-\frac{1}{2}(D_{xx}+D_{yy})|^{2}$ instead of exponentially, as it happens 
for the high-temperature regime. As a consequence, a strategy solely based on increasing the single-ion 
magnetic anisotropy might be insufficient to protct spins from these relaxation processes. Besides the 
coupling of different magnetic centres~\cite{Albold2019} the engineering of molecular vibrations represent an 
interesting possibility~\cite{Escalera-moreno2020}. First-principles simulations of spin relaxation offer a new 
perspective on this challenge as they open up the possibility to directly address the origin of the overall relaxation 
rate, as represented by the spin phonon coupling coefficients $(\partial \mathbf{D}/\partial Q_{\alpha\mathbf{q}})$. 
In order to gain an insight on this aspect of the spin dynamics we must unveil the molecular nature of the 
vibrations contributing to the spin relaxation. 

We have then analysed the effective spin-phonon coupling along the direction $\Gamma-X$ in the Brillouin zone. 
This quantity is defined as $|V_{\alpha\mathbf{q}}|^{2}=\sum_{s,t=x,y,z}(\partial D_{st}/\partial Q_{\alpha\mathbf{q}})^{2}$. 
Panels A and B of Fig.~\ref{modes} report the result of this analysis in the most relevant energy windows for both 
Orbach and Raman relaxation. Firstly, we note that $|V_{\alpha\mathbf{q}}|^{2}$ in the high-energy window is one 
order of magnitude larger that in the low-energy one. This is in agreement with other computational studies, 
where high-energy optical phonons have been found more strongly coupled to the spin with respect to the acoustic ones
and those belonging to low-energy optical branches~\cite{Lunghi2017a,Lunghi2019b}. Indeed, the largest spin-phonon 
coupling is observed for a mode occurring at $\sim 160$ cm$^{-1}$. As depicted in Fig.~\ref{modes}D, the atomic 
displacements associated with this vibration largely involves the modulation of the Co-N bond-distances and, to a 
minor degree, of all the degrees of freedom associated to the metal-ligand coordination sphere. The strong spin-phonon 
coupling of this type of motion has also been reported in both theoretical and experimental recent investigations on 
Co$^{2+}$ SIMs~\cite{Rechkemmer2016,Moseley2018,Lunghi2020}. The vibrational mode that more closely matches 
the spin excited state energy is pictured in Fig.~\ref{modes}C and it mostly involves the relative rotation of the two 
ligands with respect to the axis passing through them and the Cobalt ion. Interestingly the first vibrational state at 
the $\Gamma$ point corresponds to a rigid molecular translation and it is therefore inactive (see Fig.\ref{modes}E). 
In contrast, acoustic phonons, generally consisting of rigid molecular translations, get activated towards the Brillouin 
zone boundary as they couples to molecular rotations \cite{Lunghi2019b,Garlatti2020a} (see Fig.~\ref{modes}F).

\section*{Conclusions}

In conclusion, we have delivered a comprehensive understanding of the spin-phonon relaxation at the atomistic 
level in a prototypical Co$^{2+}$ complex with large anisotropy barrier. We have then provided an interpretation 
of both the Orbach and Raman spin-relaxation mechanisms. These results suggest that the inclusion of synthetic 
guidelines that explicitly address the interaction between spins and lattice vibrations have the potential to further 
push the limits of the spin lifetime in SIMs. In this regards, first-principles spin dynamics simulations provide a 
unique opportunity to obtain fundamental insights on the spin lifetime in magnetic molecules and drive its optimization.  

\vspace{0.2cm}
\noindent
\textbf{Acknowledgements}\\
This work has been sponsored by AMBER (grant 12/RC/2278\_P2). Computational resources were provided by the Trinity Centre for High Performance Computing (TCHPC) and the Irish Centre for High-End Computing (ICHEC). We also acknowledge the MOLSPIN COST action CA15128.\\


\end{document}